\documentclass[a4paper,twoside]{article}
\usepackage[authoryear]{natbib}
\bibpunct{(}{)}{;}{a}{}{,}
\usepackage{graphicx}
\date{} %Please leave the date blank
\newcommand{\Mo}{{\rm M}_\odot}
\newcommand{\Lo}{{\rm L}_\odot}
\title{\large\bf\flushleft A Small Step on the Long Road to Understanding the R-Stars: 
CNO Cycling in Candidate R-Star Progenitors}
\author{\parbox{\textwidth}{\flushleft
\vspace{-0.5cm}
{\it George Angelou, John Lattanzio}\\
\vspace{0.4cm}
{\small\ Centre for Stellar and Planetary Astrophysics, School of Mathematical Sciences, Monash University,
Melbourne VIC 3800, Australia
}\\
{\small\ John.Lattanzio@sci.monash.edu.au}}}
\begin{document}
%
%\twocolumn[
\begin{changemargin}{.8cm}{.5cm}
\begin{minipage}{.9\textwidth}
\vspace{-1cm}
\maketitle
%
%
%%%%%%%%%%%%%     ABSTRACT    %%%%%%%%%%%%%
%Abstract of no more than 200 words here.
\small{\bf Abstract: Recent work has proposed that a merger event between
a red-giant and a He white dwarf may be responsible for the production
of R-stars (Izzard et al, 2007). We investigate the proposed evolution
and nucleosynthesis of such a model. We simulate the hypothesized
late ignition of the core flash by increasing the neutrino losses until the
ignition occurs sufficiently far from the centre that the subsequent evolution
produces dredge-up of carbon to the extent that the post-flash 
object is a carbon star. Detailed nucleosynthesis is performed within 
this approximation, and we show that the overall properties are 
broadly consistent with the observations. Details will 
depend on the dynamics of the merger event.}

%%%%%%%%%%%%%     KEYWORDS    %%%%%%%%%%%%%
\medskip{\bf Keywords:} Write keywords here
% Please write all keywords in lower case. PASA uses the
% standard list of subject headings adopted by The Astrophysical Journal
% and available from http://www.journals.uchicago.edu/ApJ/keywords_text.html.
% Keywords are separated by em-dashes, i.e. ---

%%%%%%%%DO NOT EDIT%%%%%%%%%%%%
\medskip
\medskip
\end{minipage}
\end{changemargin}

\small
%%%%%%%%EDIT FROM HERE%%%%%%%%%%%%

\section{Introduction}
%Please see the PASA Style Guide for help with correct layout for your manuscript.
%Examples of tables and figures are given below.
Despite their discovery now dating back almost a century, a full 
explanation for the R-stars eludes us. The division
into early-R and later-R now seems to be a division into true-R 
and N (or J), respectively. We will assume this
dichotomy in what follows and pursue an explanation for the 
early-R stars. The main features to be explained are
1) they are Carbon stars. I.e. they have atmospheric \textit{n}(C) $>$
\textit{n}(O). 2) Their spectral type is otherwise K. 3) They are
enhanced in $^{12}$C, $^{13}$C, $^{14}$N, but seem to have 
solar [Fe/H], oxygen, and s-process abundances. 
4) Their luminosity (about $100\,\Lo$) identifies them as 
clump giants, that is, low mass stars burning He in their cores; and 
most peculiarly of all 5) long term studies by McClure~(2007)
failed to find any early R-stars in binary systems. Both the 
luminosity and the solar s-process abundances imply the R~stars 
have not reached the thermally pulsing AGB phase. In contrast 
are the N-stars, rich in s-process elements and with luminosities 
in excess of $2000\,\Lo$, leading to their interpretation as 
AGB stars having undergone third dredge up. The most thorough 
investigation of the composition of the R-stars was that of 
Dominy (1984).

The fact that R-stars are observed only as single stars leads 
to the notion, initially counter-intuitive, that they
must all have originated as binaries. The argument is that a single 
star would not evolve any differently to a widely
separated binary, so if R-stars are not found in binaries at all 
then they cannot exist as single stars. Hence they
must be exclusively binary in origin and their current 
singularity is assumed to be due to a merger event. Given that
approximately 20\% of normal late-type giants are binaries, 
and that none of these stars are observed in binaries,
we assume that every R-star is the product of a coalescence.

In normal low-mass single-star evolution, neutrino losses at 
high density cause a temperature inversion in
the degenerate core of stars ascending the red-giant branch. 
Eventually the triple-$\alpha$ reaction ignites at the point where the
temperature peaks, which is no longer at the centre. A strong flash 
occurs, perhaps leading to $10^9\,\Lo$ from He-burning. A
convective region develops and extends from the off-centre 
temperature maximum almost all the way to the
H-rich envelope. It seems that contact is not made between these 
two convective regions (except for the case of
very low [Fe/H]: see Fujimoto et al. 1990, 2000;
Hollowell et al. 1990;
Schlattl et al. 2001, 2002; Picardi et al. 2004;
Komiya et al. 2007). After the flash dies down, there is 
a second flash, somewhat closer to the centre
but substantially less energetic. This repeats a few times until 
the flash moves to the centre, and then central He
burning is initiated. The energy released from the explosive 
He-burning has effectively lifted the degeneracy of
the core and enables it to now burn He quiescently.

The first attempt at an explanation for the R-stars was made by Paczy\'{n}ski and
Tremaine (1977). They showed that, if the core-flash could be ignited
sufficiently far from the centre of the star, that is, at a much larger core
mass than normal, then a dredge-up episode follows the flash and carbon is
dredged to the stellar surface.  This would explain the observed $n({\rm C})
> n({\rm O})$ in the R-stars: which are thought to be core He burning stars,
and thus would be the progeny of this unusual core-flash. It remained to explain
why only a small fraction of core flashes produced such dredge-up or,
alternatively, why only a small fraction of core flashes begin at much larger
core mass than normal.

This model was the preferred explanation for R-stars 
until the discovery that they are all single stars. A
merged binary model was the basis for a recent study by 
Izzard et al. (2007) to explore merger scenarios using
binary star population synthesis. They identified possible 
formation channels that lead to an R-star outcome. The
most promising scenario was the merging of  a He white dwarf and a 
first-ascent red-giant. Typical He white dwarf masses are about
$0.15 - 0.2\,M_{\odot}$, and the red-giant mass is around $1 - 2\,\Mo$. 
The merger is hypothesised to lead to a more rapidly rotating core than normal
which then supports the core more than in the normal
case. The core flash is hence delayed and ignites at a larger core mass,
generating dredge-up in the manner found by Paczy\'{n}́ski and Tremaine (1977).

In this paper we try to take the next step in investigating this 
model, by looking at some basic nucleosynthetic
constraints. We assume that the merger event has already occurred. 
Furthermore we assume that the star has
returned to hydrostatic equilibrium, which allows us to use 
a hydrostatic stellar evolution code to model the evolution and 
nucleosynthesis. We try to force a core-flash event that is
followed by dredge-up of carbon, 
and see if the resulting abundances
are consistent with those observed in R-stars.

\section{Method}
We use the Monash version of the Mt. Stromlo stellar evolution code, 
MONSTAR (Frost and Lattanzio, 1996), for the evolutionary calculations.
 We wish to examine the 
nucleosynthesis that results from a core-flash that occurs
unusually far from the centre of the star. The hypothesis 
is that this happens as a result of the merger process, 
possibly through spinning-up the core which delays the ignition of helium.
Alternatively
during the merger itself there may be ignition of the 
core in the very outer regions.
In the absence of a code capable of calculating the merger
we have resorted to a simple artifice. The 
usual off-centre temperature maximum seen in red-giants is
due to neutrino losses in the core; since these depend mostly 
on the density they are higher in the centre, leading to
a relative cooling compared to material at a slightly 
lower density just outside the very centre. We have chosen to
increase these standard neutrino losses by an arbitrary 
factor $f_{\nu}$. This has the effect 
of cooling the core somewhat
and delaying the ignition.
It only affects the
core, however, this is the only place where the neutrino 
losses are significant. We thus expect this technique to
be sufficiently realistic for our purposes of forcing 
a more off-centre core-flash.
     
We have evolved three models of solar composition. The first was 
our standard case, with $M = 1.5\,\Mo$ and $f_{\nu} = 1$,
hereinafter M1.5NL1. The second case had $f_\nu$ increased until 
we found that the core-flash led to a dredge-up event.
This required $f_{\nu} = 30$, so this model is referred to as M1.5NL30. 
We repeated this test with a $2\,\Mo$ model and
again we required $f_{\nu} = 30$ to force a dredge-up event 
after the core flash. This model is hereinafter referred to as
M2NL30.

The models were evolved from the main sequence through to 
the giant branch and the core helium flash. We
followed the evolution to the beginning of quiescent core 
He-burning, but not beyond. This covered the dredge-up
event that followed the core flash. In the following 
sections we compare the evolution of the regular M1.5NL1
model to M1.5NL30 to see the effect of the delayed flash and 
the associated dredge-up. We also compare with the
slightly more massive M2NL30, which is also in the mass-range 
for R-star progenitors predicted by Izzard et al. (2007).
We then use our post-processing nucleosynthesis code 
MONSOON (e.g. Lattanzio et al. 1996, Lugaro et al. 2004)
to investigate the resulting nucleosynthesis and how 
this affects the surface compositions in our proposed R-stars.

Because we wish to simulate a coalesced star at a specific mass, 
we have ignored mass-loss during the evolutionary calculations. 
We are well aware of the difference between our models used 
here and the complex events that take place during the merger 
of two stars. We assume that the merged object resembles a normal giant in
structure, albeit with core and envelope masses that differ from those
that arise during normal single-star evolution. We
expect rotating cores to rapidly slow from magnetic braking.
We note, in fact, Dominy's study showed no rotational 
line broadening in his stars' line profiles.
If merged objects are the progenitors of
the R-stars, then we require their cores to still be rotating 
rapidly at the time of the core-flash, which may mean
that the mergers are required to occur only near the top of 
giant branch. Thus only a
small fraction of the mergers are likely to then experience the 
dredge-up event required to make R-stars. This
is consistent with the Izzard et al. (2007) calculations 
which found 
that if all mergers were to become R-stars then
they over-produced R-stars by a factor of ten or more. 
Thus rotational braking is a natural way to reduce the
predicted numbers to something closer to that observed.

\section{Results and Discussion}
\subsection{Structure and Evolution}
We first begin by looking at the standard evolution through 
the core flash (see for example Despain 1981, Catelan
et al. 1996). The increasing density in the core results in 
growing energy loss through neutrino processes, mostly
plasma emissions. This results in a relative cooling of the 
centre and hence an off-centre temperature maximum
develops. It is at this maximum in the temperature that 
the He ignites, under degenerate conditions,
at $m = 0.20\,\Mo$. The decoupling of temperature from 
the equation of state results in the thermal runaway
known as the flash. This peak in the He luminosity is very large, 
reaching over $10^{9}\,\Lo$ before decreasing again.
Most of the energy released goes into altering the equation 
of state (removing the degeneracy) and the core 
oscillates slightly, resulting in a second, third and fourth 
smaller flash. These are shown in Figure 1. The star then
settles down to quiescent core helium burning on the clump 
(for masses around $1-2\,\Mo$) or the horizontal branch
(for lower masses). The timescale from the major flash to quiescent 
He burning is found by all authors to be
1-2 million years (eg Despain 1981, Siess 2008), although the number 
of mini-flashes depends sensitively on the stellar mass and composition
and quite possibly the numerical details.
Figure 2 shows the extent of the convective 
regions during this phase of the evolution.
One can see clearly the convective regions generated by the 
He-burning and that the fourth such flash ends in
quiescent core helium burning in a convective core.
This model shows typical evolution through a core flash. 
The initial convective 
pulse extends close to, but does not penetrate, the
hydrogen-rich envelope. In 
particular, we note that there is no dredge-up
of carbon following  the flash.

\begin{figure}
\begin{center}
\includegraphics[scale=0.5]{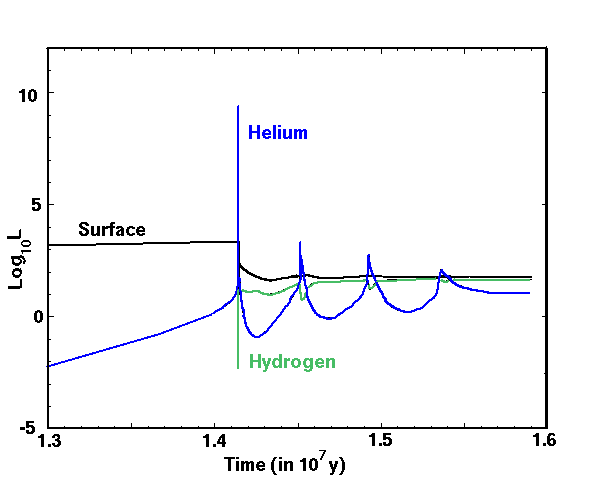}
\vspace{3mm}
\caption{The variation of the total, H and He luminosities during the core flash 
of the $M = 1.5\,\Mo$ model with normal neutrino losses. Time is measured since 
$t=2.67 \times 10^{9}\,$years.}
%\end{center}
%\end{figure}

\vspace{5mm}

%\begin{figure}
%\begin{center}
%\includegraphics[scale=.45, angle=270, trim=0mm 0mm 0mm 0mm, clip]{newimages/m15nl1conv.pdf}
\includegraphics[scale=0.5]{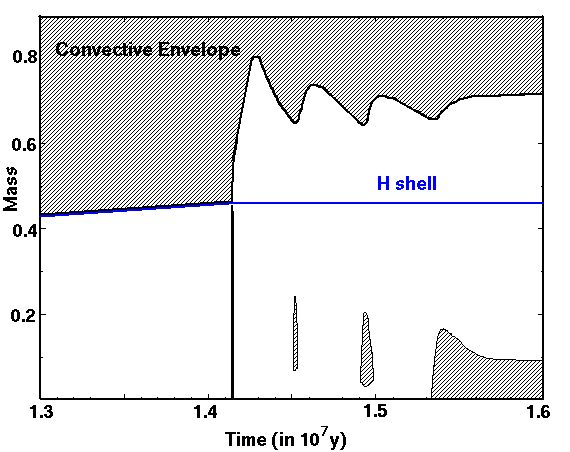}
\caption{The variation of the convective regions during the core flash 
for the standard model, as shown in Figure 1.}
\vspace{3mm}
\end{center}
\end{figure}

In contrast is our model M15NL30, with neutrino losses increased 
by a factor of 30. This model undergoes the
expected large flash but this is initiated much further out 
from the centre, due to the enhanced energy losses from
neutrinos. In this case the ignition point is at $m = 0.42\,\Mo$, 
compared to $m = 0.20\,\Mo$ in the standard case. We
note that this is followed by a series of many more small 
pulses than is seen in the standard case, and that the
time between these pulses is intiially very short, being
of order a few thousand years. This is visible
in Figure 3, and actually matches the behaviour seen in 
Paczy\'{n}ksi and Tremaine (1977; see their Figure 2).
Again, the timescale between the first pulse and the final quiescent He
burning is 1-2 million years.

The convective zones in the M1.5NL30 model are shown in 
Figure 4. We see the substantial dredge-up resulting
from the first pulse, and that there are decreasing dredge-up events 
following each of the subsequent smaller pulses,
also. The combined effect of these events is to reduce the hydrogen-exhausted 
core-mass from $0.53\,\Mo$ to $0.46\,\Mo$,
and to enhance the surface carbon content so that the star 
becomes a carbon star with n(C)/n(O) = 1.26. A
close-up of the first major dredge-up event is seen in Figure 5.

In the original calculations of the core flash there were some cases
where the convective zone at the first pulse made contact with the
hydrogen envelope, but these were determined as due to ignoring 
radiation pressure or poor numerical resolution (Despain 1981).
The first (and until this work, the only) calculation to 
show dredge-up, much like that seen
on the AGB, was Paczy\'{n}ksi and Tremaine (1977). Much like us, they
``artificially cooled'' their cores to produce a delayed flash, although
they do not give the details of how this was done. Our increased
neutrino losses are a similar artifice. In any event, we conclude that
it is the ignition of the flash much closer to the hydrogen shell that
produces an expansion and subsequent behaviour, including dredge-up, 
that is unlike other core-flash calculations and is rather
more like the dredge-up seen in  AGB stars.

Indeed, without further numerical experimentation we are unable to
definitively say whether the dredge-up is the result of the larger core 
itself or the change in the location of the
ignition point (to closer to the hydrogen
shell) or if both are needed. We have decided not to pursue this further
at present. It is indeed possible that during a merger the rapid 
accretion of matter form one core onto another may trigger the flash
further from the centre, and that the whole process is less tied to
rotation than we have proposed. In any event, the result that we are
simulating is a flash that results in dredge-up and we have succeeded in
producing that, whatever the driving mechanism.

\begin{figure}
\begin{center}
\includegraphics[scale=0.5]{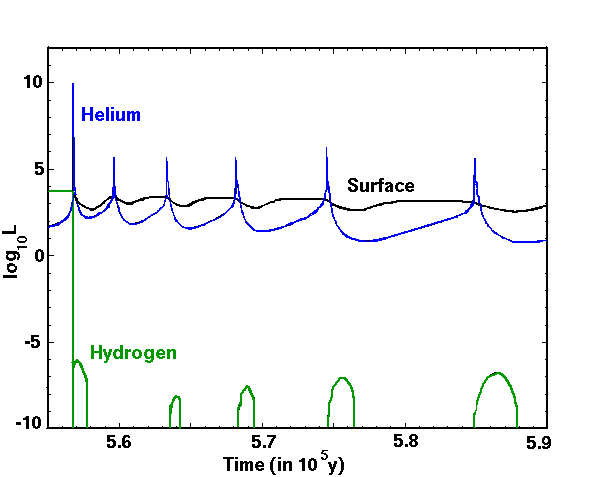}
\vspace{3mm}
\caption{The variation of the total, H and He luminosities during the core flash 
of the M1.5NL30 model with neutrino losses enhanced by a factor of 30. 
Time is measured after $t=2.67 \times 10^{9}\,$years. Note the much larger number
of small helium pulses, which occur on a much shorter timescale than in the 
standard case (Figure~1).}
%\end{center}
%\end{figure}

\vspace {5mm} 

%\begin{figure}
%\begin{center}
%\includegraphics[scale=.55, angle=270, trim=0mm 0mm 10mm 0mm, clip]{newimages/m15nl30conv.pdf}
\includegraphics[scale=0.5]{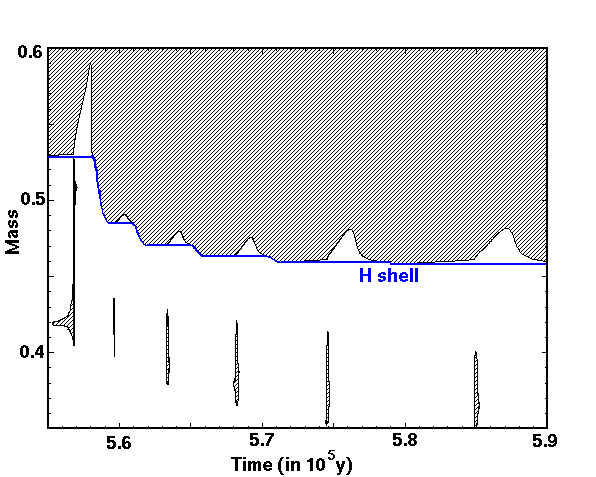}
\vspace{3mm}
\caption{The variation of the convective regions during the core flash for 
the same model as shown in Figure 3. Note the dredge-up of material following 
the major pulse. This material in enriched in carbon
and transforms the star into a carbon star. The following smaller sub-pulses also dredge
species to the envelope., but have less effect than the first event.}
\end{center}
\end{figure}

Before examining the nucleosynthesis resulting from this 
evolution, we also investigated the behaviour of a
$2\,\Mo$ model with identically enhanced neutrino 
losses. The evolution was qualitatively similar, and the convection
zones during the flash are seen in Figure 6. 
This model also became a carbon star, with $n({\rm C})/n({\rm O})= 5.6$;
this larger value indicates that more carbon is dredged to the
surface in this case.

\begin{figure}
\begin{center}
\includegraphics[scale=0.5]{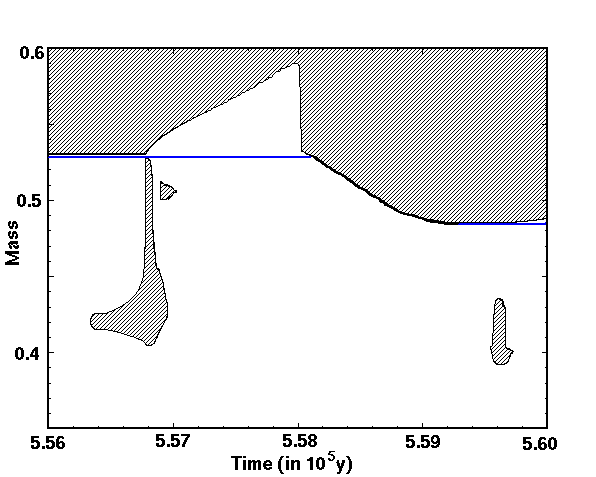}
\vspace{3mm}
\caption{Close-up of the variation of the convective regions during the first major pulse 
in model M1.5NL30, which results in a Carbon star. Once again time is 
measured after $t=2.67 \times 10^{9}\,$years.}
%\end{center}
%\end{figure}

\vspace{5mm}

%\begin{figure}
%\begin{center}
%\includegraphics[scale=.45, angle=270, trim=0mm 0mm 0mm 0mm, clip]{newimages/m2dotscon.pdf}
\includegraphics[scale=0.5]{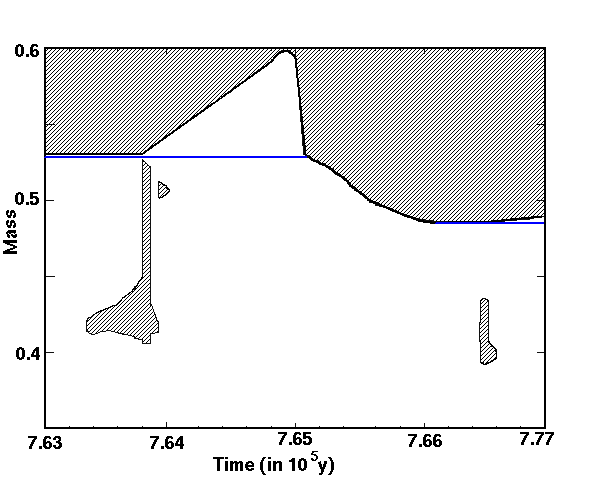}
\vspace{3mm}
\caption{Close-up of the variation of the convective regions during the core flash for 
the $2\,\Mo$ model with enhanced neutrino losses, model M2NL30,
which results in a Carbon star. Time is measured since $t=1.011 \times 10^{9}\,$years.}
\end{center}
\end{figure}

\subsection{Nucleosynthesis}
The nucleosynthesis of most interest to us is the burning 
of helium into carbon, and its subsequent mixing to the
surface. Nevertheless, we use a large network (74 species) 
to see if any other trace elements are produced which
could be used as a probe or diagnostic of the proposed mixing. 
First we discuss the CNO species, which are
the most important.

\subsubsection{CNO}
By far the most dominant effect of the altered evolution 
is that the dredge-up of primary carbon now produces
an envelope that has $n({\rm C}) > n({\rm O})$ so we form a 
carbon star as required.
Quantitative results from the simulations are 
shown in Table 1. The carbon enhancement seen in
the cases with increased neutrino loss fits reasonably well, 
certainly for the $1.5\,\Mo$ model. The
observations show substantial increases in N which we do not 
see in our models. The resulting C/O values are in fair
agreement, as is the sum C+N+O. Indeed, it is perhaps 
this latter that is most important.

\begin{table}%[p]
\begin{center}
\caption{Abundances}\label{Table 1}
\begin{tabular}{ccccc}
\hline Species$^{1}$ & Range$^{2}$ & M1.5NL1 & M1.5NL30 & M2NL30 \\
\hline
\textbf{[Fe]} & -0.40 - 0.19 & 0 & 0 & 0 \\
\textbf{[C]} & 0.12 - 0.77 & -0.12 & 0.41 & 1.07 \\
\textbf{[N]} & 0.44 - 0.82 & 0.29 & 0.28 & 0.33 \\
\textbf{[O]} & -0.53 - 0.04 & 0.0008 & -0.009 & 0.0008 \\
\textbf{C/O} & 0.9 - 3.3 & 0.36 & 1.26 & 5.6 \\
\textbf{C/N} & 1.5 - 9.3 & 1.47 & 5.2 & 21 \\
\textbf{[C+N+O]} & 0.1 - 0.4 & 0 & 0.18 & 0.63 \\
\textbf{$^{12}$C/$^{13}$C} & 4-15 & 27.5 & 99.8 & 459 \\
\hline
\end{tabular}
\medskip\\
$^1$Elemental abundances are given by number relative to the initial
(Solar) value: $[Y] = \log Y_{*} -\log Y_\odot$ \\
$^2$Range is the maximum and minimum observed values 
in the R-stars studied by Dominy (1984).\\
\end{center}
\end{table}

The observed R-stars have enhanced N, which is not 
seen in our models. Rather we have too much C.
A simple solution to this is that the C may be processed 
into N by some form of additional mixing which can transport
the enhanced C down to the H-shell for processing, much as 
happens on the first giant branch. This would also contribute
towards fixing our $^{12}{\rm C}/^{13}{\rm C}$ values which are also too large. 
All indications are that the carbon produced undergoes some CN cycling. 
We note that thermohaline 
mixing as discovered by Eggleton et al. (2006)
may be exactly what is required. Furthermore, a recent investigation 
by Cantiello et al. (2007) shows that this mixing
can operate in low mass stars during the core helium burning stage. 
This may be the way the star processes much
of the surface C into the observed N.

We have made some approximate calculations to see if burning 
various amounts of the surface carbon can
reproduce the observed abundances. The results are in Table 2. 
Here we assumed that a fraction $f$ of the
surface carbon, after dredge-up, is processed by the CNO cycle 
into N. For simplicity we assume that all this
processed C is $^{12}$C and that it appears as $^{14}$N. 
We approximate the small amount of $^{13}$C produced by adding 0.001
times the final $^{14}$N abundance to the $^{13}$C abundance 
resulting from our nucleosynthesis calculation. The results
show that for reasonable values of $f$ we are able to reproduce 
all of the observed abundances except for the carbon
isotopic ratio: the observed value is always smaller 
than our approximation. To match the observed value we must
process essentially all of the additional C through CN cycling,
which then results in overproduction of N and the
decreased C content means that the star is no longer a C star!
Similar problems exist with explaining the J stars. Note that one
effect which works in our favour is the observed deep-mixing in 
stars on the first giant branch, which reduces the carbon isotope ratio
below the value found at the first giant branch (eg Eggleton et al
2008).  This was not included in our calculations and would have
resulted in a lower ratio, perhaps closer to 14 or so, at the start of
the merger rather than the value of 21 found in these calculations.
This would go some way to reducing the discrepancy with the carbon
isotope values.

\begin{table}%[p]
\begin{center}
\caption{Approximate Results of CN Cycling}\label{Table 2}
\begin{tabular}{cccccccccccc}
\hline Species & Range & $f=0.1$ & $0.2$ & $0.3$ & $0.4$ & $0.5$ &$0.6$ & $0.7$ & $0.8$ & $0.9$ & $0.96$ \\
\hline &&&&\multicolumn{3}{c}{M1.5NL30} \\
\hline
\textbf{C/O} & 0.9 - 3.3 & 1.13 & 1.01 & 0.88 & 0.76 & 0.63  &0.50 & 0.38 & 0.25 & 0.13 & 0.05\\
\textbf{C/N} & 1.5 - 9.3 & 3.24 & 2.20 & 1.56 & 1.12 & 0.81  &0.57 & 0.38 & 0.23 & 0.10 & 0.04\\
\textbf{$^{12}$C/$^{13}$C} & 4 - 15 & 90 & 80 & 70 & 60 & 50 &40 &30 &20  &10  &4 \\
%\textbf{$^{12}$C/$^{13}$C$_\ast$} & & 19 & 17 & 15 & 13 & 11 &8.4 &6.3 &4.2  &  & \\
\textbf{[N]} & 0.44-0.82 & 0.44 & 0.55 & 0.64 & 0.72 & 0.78  &0.84 & 0.89 & 0.93 & 0.98 & 1.00\\
\hline &&&&\multicolumn{3}{c}{M2NL30} \\
%\hline Abundance & Range & f=0.1 & f=0.2 & f=0.3 & f=0.4 & f=0.5 & f=0.7 & f=0.9 & f=0.98 \\
\hline \textbf{C/O} & 0.9 - 3.3 & 5.06 & 4.50 & 3.94 & 3.37 & 2.81 & 2.25 & 1.69 & 1.12 &0.56  &0.23 \\
\textbf{C/N} & 1.5 - 9.3 & 6.74 & 3.65 & 2.29 & 1.54 & 1.05 &0.71 & 0.46 & 0.27 & 0.12 & 0.05\\
\textbf{$^{12}$C/$^{13}$C} & 4 - 15 & 411 & 364 & 317 & 270 & 224 &178 & 133 & 89  &44  &18 \\
%\textbf{$^{12}$C/$^{13}$C$_\ast$} & & 411 & 364 & 317 & 270 & 224 &0.60 & 133 & 0.80  &44  &18 \\
\textbf{[N]} & 0.44-0.82 & 0.78 & 0.99 & 1.14 & 1.24 & 1.33 &1.40 & 1.46 & 1.52 & 1.56  &1.59 \\
\hline
\end{tabular}
%\medskip\\
\end{center}
\end{table}

Note that we have assumed CN cycling and not ON cycling. The fact 
that the R-stars seem to show solar or slightly 
sub-solar oxygen may indicate
that our simple approximation is not enough. A calculation 
including the ON cycle is
required, but this would better take place within the context of a 
model for the deep-mixing itself, and is deferred
to subsequent work.

\subsubsection{Other Species}
We have included in our calculation some 74 species, 
including a small iron-peak network. With very few exceptions,
the abundance changes seen in the potential R-star models
are negligible, being below 0.1\,dex. The exceptions are the surface
abundance of $^{18}$O, and $^{22}$Ne. We see a reasonably 
large increase in the heavy oxygen isotope by a factor of
about 30, from the totally negligible $X_{18} \simeq 2\times 10^{-5}$ to the 
mostly negligible $X_{18} \simeq 5\times 10^{-4}$. 
We do not expect this to be of diagnostic assistance however.
For the unobservable $^{22}$Ne the increase is a factor of 10-20, 
to the level of $X_{22} \simeq 0.001$. There is also a small, 
but temporary, increase in the radioactive isotope $^{26}$Al.
Unfortunately, we see no changes in any species which may be used
to test the hypothesis. Our best bet lies with the CNO species.

\section{Conclusion}
The R-stars continue to resist theorists' attempts to determine their origin.
The binary merger hypothesis seems to be the best candidate
at present, but direct calculations of this stage are unavailable and
we are forced to make small steps toward validating, or otherwise, this
qualitative model. In this paper we have simulated the events that would follow
a late ignition of the core flash.
If this ignition occurs further from the centre than is normal, then we 
confirm that dredge-up of carbon may result. Our calculations show that
a substantial fraction of this carbon must then be exposed to burning via the 
CN cycles (and possibly ON). The observed low C isotope ratio remains a 
problem for the calculations shown here: the observed value indicates that 
essentially all of the added
material has been burned to equilibrium via CN (and possibly ON) cycling.
Only in that case can we match the observed $^{12}$C/$^{13}$C ratio. But
then we burn too much C into N, overproducing N and destroying
so much C that the star is no longer a carbon star.

Further advances in understanding the R-stars may require fully 3D
hydrodynamical calculations of the merger event. Such work may be possible soon
using the {\it Djehuty\/} code (e.g. Dearborn, Lattanzio and Eggleton, 2006).

\bigskip\bigskip\bigskip\bigskip

\section*{Acknowledgements} %If needed
This work was partially supported by the Australian Research Council and the Victorian Partnership for Advanced
Computing. Thanks to Ross Church and Robert Izzard for discussions and
comments. Thanks also to Lionel Siess for pointing out that a post-FDU
merger event would assist with the
Carbon isotope ratios. JL also thanks the
Australian wine industry for supporting this work, albeit indirectly. 
GA wishes to thank Carolyn Doherty, Simon Campbell and the other post grads in 
CSPA for their encouragement and support. 

\bigskip\bigskip\bigskip\bigskip

\section*{References}

\noindent Catelan, M., de Freitas Pacheco, J. A., and Horvath, J. E., 1996, {\sl Astrophys J\/}, 461, 231

\noindent Cantiello, M., Hoekstra, H., Langer, N., and Poelarends, A. J. T., 2007, 
in ``Unsolved Problems in Stellar Physics: A Conference in Honor of Douglas Gough'', 
AIP Conference Proceedings, Volume 948, pp. 73-77

\noindent Dearborn, D. S. P., Lattanzio, J. C. and Eggleton, P. P., 2006,
{\sl Astrophys. J.\/}, 639, 405

\noindent Despain, K. H., 1981, {\sl Astrophys J\/}, 251, 639

\noindent Dominy, J. F., 1984, {\sl Astrophys J Suppl Ser\/}, 55, 27

\noindent Eggleton, P. P., Dearborn, D. S. P., and Lattanzio, J. C., 2006, {\sl Science\/}, 314, 1580

\noindent Eggleton, P. P., Dearborn, D. S. P., and Lattanzio, J. C., 2008,
{\sl Astrophys. J.\/}, 677, 581

\noindent Frost, C. A., and Lattanzio, J. C., 1996, {\sl Astrophys J\/}, 473, 383

\noindent Fujimoto, M. Y., Iben, I., Jr., and Hollowell, D., 1990,
{\sl Astrophys J\/}, 349, 580

\noindent Fujimoto, M. Y., Ikeda, Y., and Iben, I., Jr., 2000,
{\sl Astrophys J Lett\/}, 529, 25

\noindent Hollowell, D., Iben, I., Jr., and Hollowell, D., 1990,
{\sl Astrophys J\/}, 351, 245

\noindent Izzard, R. G., Jeffery, C. S., and Lattanzio, J. C., 2007,
{\sl Astron Astrophys\/}, 470, 661

\noindent Komiya, Y., Suda, T. Minaguchi, H., Shigeyama, T., Aoki, W., Fujimoto, M. Y., 2007,
{\sl Astrophys J\/}, 658, 367

\noindent Lattanzio, J. C., Frost, C. A., Cannon, R. C., and Wood, P. R., 1996, {\sl Mem S A It\/}, 67, 729

\noindent Lugaro, M., Ugalde, C., Karakas, A. I., Gorres, J., Wiescher, M., Lattanzio, J. C., and Cannon, R. C., 2004,
{\sl Astrophys J\/}, 615, 934

\noindent McClure, R. D., 1997, {\sl Pub Astron Soc Pac\/}, 109, 256

\noindent Paczy\'{n}ski, B.,  and Tremaine, S. D., 1977, {\sl Astrophys J\/}, 216, 57

\noindent Picard, I., Chieffi, A., Limongi, M., Pisanti, O., Miele, G., Mangano, G., Imbriani, G., 
2004, {\sl Astrophys J\/}, 609, 1035

\noindent Schlattl, H., Cassisi, S., Salaris, M. and Weiss, A., 2001,
{\sl Astrophys J\/}, 559, 1082

\noindent Schlattl, H.,  Salaris, M., Cassisi, S., and Weiss, A., 2002,
{\sl Astron Astrophys\/}, 395, 77

\end{document}